Running head: Multi-head Watson-Crick automata
Title: Multi-head Watson-Crick automata
Authors: Kingshuk Chatterjee[1], Kumar Sankar Ray(corresponding author)[2]

Affiliations:
[1]Electronics and Communication Sciences Unit, Indian Statistical Institute, Kolkata-108

[2]Professor, Electronics and Communication Sciences Unit, Indian Statistical Institute, Kolkata-108

Address: Electronics and Communication Sciences Unit, Indian Statistical Institute, Kolkata-108.

Telephone Number: +918981074174

Fax Number:033-25776680

Email:ksray@isical.ac.in

# Multi-head Watson-Crick automata

*Kingshuk Chatterjee,[1] Kumar Sankar Ray[2]*

*Electronics and Communication Science unit, ISI, Kolkata.*

[1]kingshukchaterjee@gmail.com [2]ksray@isical.ac.in

*Abstract:* Inspired by multi-head finite automata and Watson-Crick automata in this paper, we introduce new structure namely multi-head Watson-Crick automata where we replace the single tape of multi-head finite automaton by a DNA double strand. The content of the second tape is determined using a complementarity relation similar to Watson-Crick complementarity relation. We establish the superiority of our model over multi-head finite automata and also show that both the deterministic and non-deterministic variant of the model can accept non-regular unary languages. We also compare our model with parallel communicating Watson-Crick automata systems and prove that both of them have the same computational power.

## I. INTRODUCTION

Piatkowski[1], Rabin et al[2] and Rosenberg[3] introduced multi-head finite automata in the sixties. Holzer[4] worked on 2 way multi-head finite automata. Yao et al.[5] established that k+1 heads are better than k and that there is a language recognised by a non-deterministic multi-head finite automata but not by any deterministic multi-head finite automata. Kutylowski[6] also discussed about the head hierarchy of deterministic multi-head finite automata.

Păun et.al.[7] introduced Watson-Crick automata (Watson-Crick automata are finite automata having two independent heads working on double strands where the characters on the corresponding positions of the two strands are connected by a complementarity relation similar to the Watson-Crick complementarity relation. The movement of the heads although independent of each other is controlled by a single state). Czeizler et.al. [8] introduced the deterministic variants of Watson-Crick automata. Work on state complexity of Watson-Crick automata is discussed in [9] and [10]. Parallel Communicating Watson-Crick automata systems (PCWKS) were introduced in [11] and further investigated in [12].

Inspired by multi-head finite automata and Watson-Crick automata, we introduce a new structure namely multi-head Watson-Crick automata. This model is same as the multi-head finite automata model but the reading tape is replaced by a double strand similar to the Watson-Crick model where the characters on the corresponding positions of the two strands are connected by a complementarity relation similar to the Watson-Crick complementarity relation. There are many independent heads on both the upper and the lower strands.

The main focus of this paper are as follows:

1)To formally define a multi-head finite automata model where the input tape is replaced by a DNA double strand.

2)To establish that both the deterministic and non-deterministic variants of the above mentioned model are computationally more powerful than there single tape counterparts.

3)To show that both the deterministic and non deterministic variant of the model can accept some non-regular unary languages.

4)Finally to show that the computational power of multi-head Watson-Crick automata and parallel communicating automata are same.

## II. BASIC TERMINOLOGY

The symbol V denotes a finite alphabet. The set of all finite words over V is denoted by $V^*$, which includes the empty word λ. The symbol $V^+ = V^* - \{\lambda\}$ denotes the set of all non-empty words over the alphabet V. For w ∈ $V^*$, the length of w is denoted by |w|. Let u∈ $V^*$ and v ∈$V^*$ be two words and if there is some word x ∈ $V^*$, such that v=ux, then u is a prefix of v, denoted by u ≤ v. Two words, u and v are prefix comparable denoted by u~$_p$v if u is a prefix of v or vice versa.

### Watson-Crick automata

A Watson-Crick automaton is a 6-tuple of the form M=(V,ρ,Q,$q_0$,F,δ) where V is an alphabet set, set of states is denoted by Q, ρ ⊆ V×V is the complementarity relation similar to Watson-Crick complementarity relation, $q_0$ is the initial state and F⊆Q is the set of final states. The function δ contains a finite number of transition rules of the form $q \binom{w_1}{w_2} \rightarrow q'$, which denotes that the machine in state q parses $w_1$ in upper strand and $w_2$ in lower strand and goes to state q' where $w_1, w_2 \in V^*$. The symbol $\begin{bmatrix} w_1 \\ w_2 \end{bmatrix}$ is different from $\binom{w_1}{w_2}$. While $\binom{w_1}{w_2}$ is just a pair of strings written in that form instead of ($w_1,w_2$), the symbol $\begin{bmatrix} w_1 \\ w_2 \end{bmatrix}$ denotes that the two strands are of same length i.e. |$w_1$|=|$w_2$| and the corresponding symbols in two strands are complementarity in the sense given by the relation ρ. The symbol $\begin{bmatrix} V \\ V \end{bmatrix}_\rho = \{ \begin{bmatrix} a \\ b \end{bmatrix} \mid a, b \in V, (a, b) \in \rho \}$ and $WK_\rho(V) = \begin{bmatrix} V \\ V \end{bmatrix}_\rho^*$ denotes the Watson-Crick domain associated with V and ρ.

A transition in a Watson-Crick finite automaton can be defined as follows:

For $\binom{x_1}{x_2}, \binom{u_1}{u_2}, \binom{w_1}{w_2} \in \binom{V^*}{V^*}$ such that $\begin{bmatrix} x_1 u_1 w_1 \\ x_2 u_2 w_2 \end{bmatrix} \in WK_\rho(V)$ and $q, q' \in Q$, $\binom{x_1}{x_2} q \binom{u_1}{u_2} \binom{w_1}{w_2} \Rightarrow \binom{x_1}{x_2} \binom{u_1}{u_2} q' \binom{w_1}{w_2}$ iff there is transition rule $q\binom{u_1}{u_2} \rightarrow q'$ in $\delta$ and $\stackrel{*}{\Rightarrow}$ denotes the transitive and reflexive closure of $\Rightarrow$. The language accepted by a Watson-Crick automaton M is L(M)={$w_1 \in V^* | q_0 \begin{bmatrix} w_1 \\ w_2 \end{bmatrix} \stackrel{*}{\Rightarrow} q \begin{bmatrix} \lambda \\ \lambda \end{bmatrix}$, with $q \in F$, $w_2 \in V^*$, $\begin{bmatrix} w_1 \\ w_2 \end{bmatrix} \in WK_\rho(V)$}.

## Subclasses of Non-deterministic Watson-Crick Automata

Depending on the type of states and transition rules there are four types or subclasses of Watson-Crick automata. A Watson-Crick automaton M=(V,ρ,Q,q₀,F, δ) is

1) stateless( NWK ): If it has only one state, i.e. Q=F={ $q_0$ };
2) all-final( FWK ): If all the states are final, i.e. Q=F;
3) simple( SWK ): If at each step the automaton reads either from the upper strand or from the lower strand, i.e. for any transition rule $q\binom{w_1}{w_2} \rightarrow q'$, either $w_1 = \lambda$ or $w_2 = \lambda$;
4) 1-limlited( 1-limited WK ): If for any transition rule $q\binom{w_1}{w_2} \rightarrow q'$, we have $|w_1 w_2| = 1$.

## Deterministic Watson-Crick Automata And Their Subclasses

The notion of determinism in Watson-Crick automata and a discussion on its complexity were first considered in [8]. In [8] different notions of determinism were suggested as follows:

1) weakly deterministic Watson-Crick automata(WDWK): Watson-Crick automaton is weakly deterministic if in every configuration that can occur in some computation of the automaton, there is a unique possibility to continue the computation, i.e. at every step of the automaton there is at most one way to carry on the computation.
2) deterministic Watson-Crick automata(DWK): deterministic Watson-Crick automaton is Watson-Crick automaton for which if there are two transition rules of the form $q\binom{u}{v} \rightarrow q'$ and $q\binom{u'}{v'} \rightarrow q''$ then $u \not\prec_p u'$ or $v \not\prec_p v'$.
3) strongly deterministic Watson-Crick automata(SDWK): strongly deterministic Watson-Crick automaton is a deterministic Watson-Crick automaton where the Watson-Crick complementarity relation is injective.

Similar to non-deterministic Watson-Crick automata, deterministic Watson-Crick automata can be stateless (NDWK), all final (FDWK), simple (SiDWK) and 1-limited (1-limited DWK).

## Parallel communicating Watson-Crick automata system

A parallel communicating Watson-Crick automata system of degree *n*, denoted by PCWK(n), is a *(n + 3)*-tuple

A = *(V, ρ, A₁, A₂, . . . , Aₙ, K)*,

where

- *V* is the input alphabet;
- *ρ* is the complementarity relation;
- $A_i = (V, \rho, Q_i, q_i, F_i, \delta_i)$, $1 \leq i \leq n$, are Watson-Crick finite automata, where the sets $Q_i$ are not necessarily disjoint;
- $K = \{K_1, K_2, \ldots, K_n\} \subseteq \bigcup_{i=1}^{n} Q_i$ is the set of query states.

The automata $A_1, A_2, \ldots, A_n$ are called the *components* of the system A. Note that any Watson-Crick finite automaton is a parallel communicating Watson-Crick automata system of degree 1.

A configuration of a parallel communicating Watson-Crick automata system is a 2*n*-tuple ($s_1$, $\binom{u_1}{v_1}$, $s_2$, $\binom{u_2}{v_2}$, . . . , $s_n$, $\binom{u_n}{v_n}$) where $s_i$ is the current state of the component *i* and $\binom{u_i}{v_i}$ is the part of the input word which has not been read yet by the component *i*, for all $1 \leq i \leq n$. We define a binary relation ⊢ on the set of all configurations by setting

($s_1$, $\binom{u_1}{v_1}$, $s_2$, $\binom{u_2}{v_2}$, . . . , $s_n$, $\binom{u_n}{v_n}$) ⊢ ($r_1$, $\binom{u_1'}{v_1'}$, $r_2$, $\binom{u_2'}{v_2'}$, . . . , $r_n$, $\binom{u_n'}{v_n'}$)

if and only if one of the following two conditions holds:

- K ∩ {$s_1, s_2, \ldots, s_n$} = ∅, $\binom{u_i}{v_i} = \binom{x_i}{y_i} \binom{u_i'}{v_i'}$, and $r_i \in \delta_i (s_i, \binom{x_i}{y_i})$, $1 \leq i \leq n$;

- for all $1 \leq i \leq n$ such that $s_i = K_{j_i}$ and $s_{j_i} \notin K$ we have $r_i = s_{j_i}$, whereas for all the other $1 \leq \ell \leq n$ we have $r_\ell = s_\ell$. In this case $\binom{u'_i}{v'_i} = \binom{u_i}{v_i}$, for all $1 \leq i \leq n$.

If we denote by $\vdash^*$ the reflexive and transitive closure of $\vdash$, then the language recognized by a PCWKS is defined as:

$L(A) = \{w_1 \in V^* \mid (q_1, \begin{bmatrix}w_1\\w_2\end{bmatrix}, q_2, \begin{bmatrix}w_1\\w_2\end{bmatrix}, \ldots, q_n, \begin{bmatrix}w_1\\w_2\end{bmatrix}) \vdash^* (s_1, \begin{bmatrix}\lambda\\\lambda\end{bmatrix}, s_2, \begin{bmatrix}\lambda\\\lambda\end{bmatrix}, \ldots, s_n, \begin{bmatrix}\lambda\\\lambda\end{bmatrix}), s_i \in F_i, 1 \leq i \leq n\}$.

Intuitively, the language accepted by such a system consists of all words $w_1$ such that in every component we reach a final state after reading all input $\begin{bmatrix}w_1\\w_2\end{bmatrix}$. Moreover, if one of the components stops before the others, the system halts and rejects the input. The above definition of parallel communicating Watson-Crick automata is in [11].

## Deterministic Parallel communicating Watson-Crick automata system

The notion of determinism in deterministic parallel communicating Watson-Crick automata is as follows.

1) weakly deterministic parallel communicating Watson-Crick automata system(WDPCWKS): a parallel communicating Watson-Crick automata system is weakly deterministic if every component in the system is a weakly deterministic Watson-Crick automaton.

2) deterministic parallel communicating Watson-Crick automata system(DPCWKS): a parallel communicating Watson-Crick automata system is deterministic if every component in the system is a deterministic Watson-Crick automaton.

3) strongly deterministic parallel communicating Watson-Crick automata system(SDPCWKS): a parallel communicating Watson-Crick automata system is strongly deterministic if every component in the system is a strongly deterministic Watson-Crick automaton.

## Multi-head Finite automata

Non-deterministic multi-head finite automaton is a 6 tuple $M=(k,V,Q,q_0,F,\delta)$ where V is an alphabet set, set of states is denoted by Q, the symbol $q_0$ is the initial state and $F \subseteq Q$ is the set of final states. The number of heads is denoted by k. The input is of the form $w \in V^*$. The function $\delta$ contains a finite number of transitions of the form $\delta(q_i,(a_1,a_2,...,a_k))=q_j$ where $a_i \in V \cup \{\lambda\}$, $q_i,q_j \in Q$ in M, which denotes that the machine in state $q_i$ parses $a_1,a_2,.....,a_k$ using the k heads and goes to state $q_j$.

For $x_i,u_i,w_i \in V^*$ where i=1 to k such that $x_iu_iw_i=w \in V^*$ where i=1 to k and $q, q' \in Q$, $(x_1,x_2,...,x_k)q(u_1w_1,u_2w_2,...,u_kw_k) \Rightarrow (x_1u_1,x_2u_2,...,x_ku_k)q'(w_1,w_2,...,w_k)$ iff there is a transition $\delta(q,(u_1,u_2,...,u_k))=q'$ in $\delta$ and $\stackrel{*}{\Rightarrow}$ denotes the transitive and reflexive closure of $\Rightarrow$.

The string accepted by a multi-head finite automaton M is the string on its input tape when all the heads have completely read the input and the automaton is in its final state. i.e. $L(M)=\{w \in V^* \mid (\lambda,\lambda,...,\lambda)q_0(w,w,...,w) \stackrel{*}{\Rightarrow} (w,w,...,w)q(\lambda,\lambda,...,\lambda)$ with $q \in F\}$.

Deterministic multi-head automaton has the same structure as the non-deterministic variant with an additional restriction that if there are transition rules of the form $\delta(q_i,(a_1,a_2,...,a_k))=q_j$ and $\delta(q_i,(c_1,c_2,...,c_k))=q_k$ then there must atleast be one pair of $a_i,c_i$ i=1 to k such that $a_i \not\sim_p c_i$.

### III. MULTI-HEAD WATSON-CRICK AUTOMATA

Multi-head Watson-Crick automaton is derived from multi-head finite automata model by replacing the input tape by a DNA double strand and introducing multiple heads, which are independent of each other on each strand.

**Definition 1**: Multi-head Watson-Crick automaton is a 8 tuple $M=(V,\rho,Q,q_0,F,\delta,k_1,k_2)$ where V is an alphabet set, set of states is denoted by Q, the complementarity relation $\rho \subseteq V \times V$ is similar to Watson-Crick complementarity relation, $q_0$ is the initial state and $F \subseteq Q$ is the set of final states. The symbol $k_1$ denotes the number of heads on the upper strand and $k_2$ denotes the number of heads on the lower strand. The input is of the form $\begin{bmatrix}w_1\\w_2\end{bmatrix}$ where $\begin{bmatrix}w_1\\w_2\end{bmatrix} \in WK_\rho(V)$. The function $\delta$ contains a finite number of transitions of the form $\delta(q_i,(a_1,a_2,..., a_{k_1}|b_1,b_2,..., b_{k_2}))=q_j$ where $a_m, b_n \in V \cup \{\lambda\}$, $1<=m<=k_1$, $1<=n<=k_2$ and $q_i,q_j \in Q$, which denotes that the machine in state $q_i$ parses $a_1, a_2,..., a_{k_1}$ using the $k_1$ heads in upper strand and $b_1, b_2,..., b_{k_2}$ using the $k_2$ heads in lower strand and goes to state $q_j$.

For $t_i,u_i,v_i \in V^*$ such that $t_iu_iv_i=w_1 \in V^*$ where i=1 to $k_1$ and $x_j,y_j,z_j \in V^*$ such that $x_jy_jz_j=w_2 \in V^*$ where j=1 to $k_2$ and $\begin{bmatrix}w_1\\w_2\end{bmatrix} \in WK_\rho(V)$, the symbols $q, q' \in Q$, $(t_1,t_2,...,t_{k_1}|x_1,x_2,...,x_{k_2})q(u_1v_1,u_2v_2,...,u_{k_1}v_{k_1}|y_1z_1,y_2z_2,...,y_{k_2}z_{k_2}) \Rightarrow (t_1u_1,t_2u_2,...,t_{k_1}u_{k_1}|x_1y_1,x_2y_2,...,x_{k_2}y_{k_2})q'(v_1,v_2,...,v_{k_1}|z_1,z_2,...,z_{k_2})$ iff there is a transition $\delta(q,(u_1,u_2,...,u_{k_1}| y_1,y_2,...,y_{k_2}))=q'$ in $\delta$ and $\stackrel{*}{\Rightarrow}$ denotes the transitive and reflexive closure of $\Rightarrow$.

The string accepted by a multi-head Watson-Crick automaton M is the string on the upper strand when all the heads (both on the upper and lower strands) have completely read the input and the automaton is in its final state. i.e. L(M)={$w_1 \in V^*$|$(\lambda,\lambda,...,\lambda|\lambda,\lambda,...,\lambda)q_0(w_1,w_1,...,w_1|w_2,w_2,...,w_2) \stackrel{*}{\Rightarrow} (w_1,w_1,...,w_1|w_2,w_2,...,w_2)q(\lambda,\lambda,...,\lambda|\lambda,\lambda,...,\lambda)$ with $q \in F$, $w_2 \in V^*$, $\begin{bmatrix}w_1\\w_2\end{bmatrix} \in WK_\rho(V)$} where $w_1$ is read by the $k_1$ heads on upper strand and $w_2$ is read by the $k_2$ heads on lower strand.

**Deterministic multi-head Watson-Crick automaton** has the same structure as the non-deterministic variant with an additional restriction that if there are transition rules of the form $\delta(q_i,(a_1,a_2,...,a_{k_1}|b_1,b_2,...,b_{k_2}))=q_j$ and $\delta(q_i,(c_1,c_2,...,c_{k_1}|d_1,d_2,...,d_{k_2}))=q_k$ then there must atleast be one pair of $a_i,c_i$ i=1 to $k_1$ such that $a_i \not\sim_p c_i$ or one pair of $b_j,d_j$ j=1 to $k_2$ such that $b_j \not\sim_p d_j$.

A deterministic multi-head Watson-Crick automaton is called a **strongly deterministic multi-head Watson-Crick automaton** if the complementarity relation is injective.

## IV. COMPUTATIONAL COMPLEXITY OF NON-DETERMINISTIC MULTI-HEAD WATSON-CRICK AUTOMATA

In this Section, we discuss the computational complexity of non-deterministic multi-head Watson-Crick automata.

**Theorem 1:** non-deterministic multi-head Watson-Crick automaton with 1+2 heads can accept the non-regular unary language L={$a^n$, where $n = \sum_{k=0}^{l} 2^k$, $l \in N$}.

Proof: A non-deterministic multi-head Watson-Crick automaton $M=(V,\rho,Q,q_0,F,\delta,k_1,k_2)$ accepts the unary non-regular language L={$a^n$, where $n = \sum_{k=0}^{l} 2^k$, $l \in N$} where V={a,b,c}, $\rho$ = {(a,b), (a,c)}, Q={$q_0, q_1, q_2, q_3, q_4, q_5, q_f$}, F={$q_f$}, $k_1=1$, $k_2=2$.

The transitions of M are as follows:
$\delta(q_0,(\lambda|\lambda,b))=q_1$, $\delta(q_1,(a|b,c))=q_2$, $\delta(q_2,(\lambda|\lambda,c))=q_3$, $\delta(q_3,(a|b,c))=q_2$, $\delta(q_3,(a|c,b))=q_4$, $\delta(q_4,(\lambda|\lambda,b))=q_5$, $\delta(q_5,(a|c,b))=q_4$, $\delta(q_5,(a|b,c))=q_2$, $\delta(q_2,(\lambda|\lambda,c))=q_f$, $\delta(q_3,(\lambda|\lambda,b))=q_f$, $\delta(q_f,(a|b,\lambda))=q_f$, $\delta(q_f,(a|c,\lambda))=q_f$.

The multi-head Watson-Crick automaton M works in the following manner:

At first the second lower head reads a 'b', this ensures that the string in the lower strand of M begins with 'b'. After that every time the first lower head reads a 'b/c' the second lower head reads two 'c/b' and the upper head reads an 'a'.

Thus the automaton described above accepts all strings composed of 'a' which has as complementarity strings those strings which have the following properties.
1) alternate blocks of 'b' and 'c'.
2) the first block begins with 'b' and length of first block is 1.
3) Length of all blocks are twice that of their previous blocks.

Consider a string w in L, one of the many complementarity strings of w will have the above stated property, thus M accepts w. E.g. w=aaaaaaaa∈L one of its many complementarity strings is bccbbbb.

Now, for a string not in L, none of its complementarity strings can have the above stated properties, thus M reject w.

**Corollary 1:** A non-deterministic multi-head Watson-Crick automata with 1+2 heads accepts a language not accepted by any non-determinitic multi-head finite automata.

Proof: From Theorem 1, we know that a non-deterministic multi-head Watson-Crick automata with 1+2 heads can accept the non-regular unary language L={$a^n$, where $n = \sum_{k=0}^{l} 2^k$, $l \in N$} and we also know that non-deterministic multi-head finite automata can accept only regular unary languages[12]. Thus, non-deterministic multi-head Watson-Crick automaton with 1+2 heads accepts a language not accepted by any non-deterministic multi-head finite automata.

**Lemma 1:** For every non-deterministic multi-head Watson-Crick automata with $k_1$=m and $k_2$=1 heads there is a non-deterministic multi-head automata with k= m+1 heads.

Proof: Given a non-deterministic Watson-Crick automaton $M=(V,\rho,Q,q_0,F,\delta,m,1)$, we can obtain a non-deterministic multi-head finite automaton $M'=(m+1,V,Q,q_0,F,\delta')$ where $\delta'$ is formed from $\delta$ in the following manner for transitions of the form $\delta(q_i,(a_1,a_2,....,a_m|\lambda))=q_j$ where $a_1,a_2,....,a_m \in V \cup \{\lambda\}$ we introduce $\delta(q_i,(a_1,a_2,....,a_m,\lambda))=q_j$ in $\delta'$. For transitions of the form $\delta(q_i,(a_1,a_2,....,a_m|a_{m+1}))=q_j$ where $a_1,a_2,....,a_m \in V \cup \{\lambda\}$, $a_{m+1} \in V$, we introduce transitions of the form $\delta(q_i,a_1,a_2,....,a_m,a)=q_j$ in $\delta'$ where $\rho^{-1}(a_{m+1})=a$ and 'a' must be present in the upper strand in the corresponding position as contents of the lower strand is obtained by applying the complementarity relation $\rho$ to each position of the upper strand. The non-deterministic multi-head finite automaton non-deterministically guesses the complementarity element in the lower head for 'a' and executes that particular transition. When the number of heads in the lower strand becomes more than one then this method does not work as there is no way to ensure that the non-deterministic finite automata guesses the same complementarity element for 'a' every time a lower head reads that particular position.

**Lemma 2:** For every non-deterministic multi-head finite automaton with m+1 heads there is a non-deterministic multi-head Watson-Crick automaton with $k_1=m$ and $k_2=1$ and identity complementarity relation which accepts the same language.

Proof: The proof of this Theorem is due to the fact that the complemenatrity relation is the the identity relation therefore the content of the lower strand is same as the upper strand. Therefore non-deterministic multi-head Watson-Crick automaton with $k_1=m$ and $k_2=1$ and identity complementarity relation just simulates the moves of non-deterministic multi-head finite automaton with m heads.

**Theorem 2:** Non-deterministic multi-head Watson-Crick automata with $k_2=1$ and non-deterministic multi-head finite automata have the same computational power.

Proof: From Lemma 1, we see that for every non-deterministic multi-head Watson-Crick automaton with $k_2=1$ there exists a non-deterministic multi-head finite automaton which accepts the same language and from Lemma 2, we get that for every non-deterministic multi-head finite automaton there is a non-deterministic multi-head Watson-Crick automaton with $k_2=1$ which accepts the same language L. Thus we can say that non-deterministic multi-head Watson-Crick automata with $k_2=1$ and non-deterministic multi-head finite automata have the same computational power.

**Corollary 2:** Non-deterministic multi-head Watson-Crick automata with $k_2=1$ and can accept only regular unary languages.

Proof: We know that non-deterministic multi-head finite automata can accept only regular unary languages [12] and from Theorem 2 we see that non-deterministic multi-head Watson-Crick automata with $k_2=1$ and non-deterministic multi-head finite automata have the same computational power. Hence we conclude that non-deterministic multi-head Watson-Crick automata with $k_2=1$ and can accept only regular unary languages.

**Theorem 3:** Set of languages accepted by non-deterministic multi-head finite automata is a proper subset of set of languages accepted by non-deterministic multi-head Watson-Crick automata.

Proof: From Lemma 2, we know that for every non-deterministic multi-head finite automaton with m heads there is a non-deterministic multi-head Watson-Crick automaton with $k_1=m-1$ and $k_2=1$ and identity complementarity relation which accepts the same language and from corollary 1 we see that there exists a language accepted by non-deterministic multi-head Watson-Crick automata with 1+2 heads which is not accepted by any non-determinitic multi-head finite automata, which proves the above Theorem.

**Lemma 3:** For every parallel communicating Watson-Crick automata system $A = (V, \rho, A_1, A_2, \ldots, A_n, K)$, with n components (where transitions of each component are of the form $q\binom{w_1}{w_2} \to q'$ and $|w_1 w_2| <= 1$) there is a non-deterministic multi-head Watson-Crick automaton with n+n heads which accept the same language.

Proof: The proof of the above theorem is similar to the proof in [13] for parallel communicating finite automata system and multi-head finite automata.

The proof is as follows:

Let $A = (V, \rho, A_1, A_2, \ldots, A_n, K)$, be a parallel communicating Watson-Crick automata system with n components having transitions of the form $q\binom{w_1}{w_2} \to q'$ and $|w_1 w_2| <= 1$ in its components where $A_i=(V,\rho,Q_i,q_{0i},F_i,\delta_i)$, $1<=i<=n$. The n+n head non-deterministic multi-head Watson-Crick automaton accepting the same language as the parallel communicating Watson-Crick automata system is obtained as follows.

$M=(V,\rho,(Q_1 \cup K) \times (Q_2 \cup K) \times \ldots \times (Q_n \cup K), q_0, F_1 \times F_2 \times \ldots \times F_n, \delta, n, n)$ where

$\delta((s_1,s_2,\ldots,s_n),(a_1,a_2,\ldots,a_n|b_1,b_2,\ldots,b_n))=\{(p_1,p_2,\ldots,p_n)|p_i \in \delta_i(s_i, \binom{a_i}{b_i}), a_i,b_i \in V^*, |a_i b_i|<=1, 1<=i<=n\}$, iff $\{s_1,s_2,\ldots,s_n\} \cap K = \emptyset$,

$\delta((s_1,s_2,\ldots,s_n),(\lambda,\lambda,\ldots,\lambda|\lambda,\lambda,\ldots,\lambda))= (p_1,p_2,\ldots,p_n)$

$$\text{where } p_i = \begin{cases} s_{j_i} \notin K, \text{if } s_i = K_{j_i}, \\ q_i, \text{if there exist j such that } s_j = K_i \\ s_i, \text{otherwise} \end{cases}$$

for all $1<=i<=n$

Each current state of the multi-head Watson-Crick automata keep track of all current states of the automata system.

From the construction of multi-head Watson-Crick automaton M it is evident that it accepts the same language as the parallel communicating system A.

**Lemma 4:** Every parallel communicating Watson-Crick automata System is equivalent with a system where the components have transitions of the form $q\binom{w_1}{w_2} \to q'$ and $|w_1 w_2| <= 1$ and the equivalent system has the same number of components as the parallel communicating Watson-Crick automata System.

Czeizler et. al.[6] proved Lemma 4.

**Theorem 4:** For every parallel communicating Watson-Crick automata system $A = (V, \rho, A_1, A_2, \ldots, A_n, K)$, with n components there is a non-deterministic multi-head Watson-Crick automaton with n+n heads which accept the same language.

Proof: From Lemma 4, we know that for every parallel communicating Watson-Crick automata System is equivalent with a system where the components have transitions of the form $q\binom{w_1}{w_2} \to q'$ and $|w_1 w_2| \leq 1$ and the equivalent system has the same number of components as the parallel communicating Watson-Crick automata System and from Lemma 3 we see that for every parallel communicating Watson-Crick automata system $A = (V, \rho, A_1, A_2, \ldots, A_n, K)$, with n components (where transitions of each component are of the form $q\binom{w_1}{w_2} \to q'$ and $|w_1 w_2| \leq 1$) there is a non-deterministic multi-head Watson-Crick automaton with n+n heads which accept the same language. Thus, from Lemma 3 and 4 we conclude that for every parallel communicating Watson-Crick automata system $A = (V, \rho, A_1, A_2, \ldots, A_n, K)$, with n components there is a non-deterministic multi-head Watson-Crick automaton with n+n heads which accept the same language.

**Theorem 5:** For every multi-head Watson-Crick automata with $k_1+k_2$ heads there is a parallel communicating Watson-Crick automata system with $\max(k_1,k_2)$ components which accepts the same language.

Proof: The proof of the above Theorem is similar to the proof in [13] for parallel communicating finite automata system and multi-head finite automata.

Here we consider for simplicity $k_1=k_2=n$ but the construction method described below will work even when $k_1 \neq k_2$ we will just have to take the number of components to be $\max(k_1,k_2)$. In case $k_1 \neq k_2$, the extra heads on the parallel communicating Watson-Crick automaton will just read its symbol every time the other head in the same component reads a symbol.

Let $M=(V,\rho,Q,q_0,F,\delta,n,n)$ be a multi-head Watson-Crick automata. We construct the parallel communicating Watson-Crick automata $A = (V, \rho, A_1, A_2, \ldots, A_n, K)$ in the following manner:

$A_i = (V,\rho,Q_i,q_0,F,\delta_i)$, where

$Q_i = K \cup Q \cup (Q \times (V \cup \{\lambda\})^{i-1} \times (V \cup \{\lambda\})^{i-1}) \cup (Q \times (V \cup \{\lambda\})^i \times (V \cup \{\lambda\})^i) \cup X_i \cup Y_i$

with

$X_i = \begin{cases} \emptyset, i \leq 2 \\ \{p_j|, 1 \leq i \leq i-2\}, i > 2, \end{cases}$

$Y_i = \begin{cases} \emptyset, i = n \\ \{s_j | i+1 \leq j \leq n\}, i < n, \end{cases}$

For a transition $\delta(q_i,(a_1,a_2,\ldots,a_n|b_1,b_2,\ldots,b_n)) = q_j$ $a_m, b_p \in V \cup \{\lambda\}$, $1 \leq m \leq n$, $1 \leq p \leq n$ and $q_i, q_j \in Q$

The transitions introduced in the component $A_i$ of A i.e. in $\delta_i$ ($1 \leq i \leq n$) are as follows:

All the components begin in state $q_i$, all the components are in waiting except the first component. The first component reads input from the upper and lower strands using two independent heads based on $a_1$ and $b_1$ read by the multi-head Watson-Crick automaton and stores the symbol in its current state. The control then switches to component two, and the current state of component one is passed to component two. All the other components are still in waiting.

i=1: $\delta_i(q_i, \binom{a_1}{b_1}) = (q_i, a_1, b_1)$,

$\delta_i((q_i, a_1, b_1), \binom{\lambda}{\lambda}) = s_2$,

$\delta_i(s_j, \binom{\lambda}{\lambda}) = s_{j+1}$, $2 \leq j \leq n-1$,

$\delta_i(s_n, \binom{\lambda}{\lambda}) = K_n$

Components 2 to n-1 behave in the same manner as component one that is they read the symbols according to transition of the multi-head Watson-Crick automaton the $i^{th}$ components head reads the symbol $a_i$ and $b_i$ stores the symbol read along with the symbol read information it received from the component before it in its current state. It switches the control to its next component and also passes its current state to the next state.

i=2: $\delta_i(q_i, \binom{\lambda}{\lambda}) = K_1$

$\delta_i((q_i, a_1, b_1), \binom{a_2}{b_2}) = (q_i, a_1, a_2, b_1, b_2)$,

$\delta_i((q_i, a_1, a_2, b_1, b_2), \binom{\lambda}{\lambda}) = s_3$,

$\delta_i(s_j, \binom{\lambda}{\lambda}) = s_{j+1}$, $3 \leq j \leq n-1$,

$\delta_i(s_n, \binom{\lambda}{\lambda}) = K_n$

$i=2<i<n$  $\delta_i(q_i, \binom{\lambda}{\lambda})=p_1$

$\delta_i(p_j, \binom{\lambda}{\lambda})=p_{j+1}$, $1<=j<=i-3$

$\delta_i(p_{i-2}, \binom{\lambda}{\lambda})=K_{i-1}$

$\delta_i((q_i, a_1,a_2,.....,a_{i-1},b_1,b_2,.....,b_{i-1}), \binom{a_i}{b_i})=(q_i, a_1,a_2,.....,a_i,b_1,b_2,.....,b_i)$,

$\delta_i((q_i, a_1,a_2,.....,a_i,b_1,b_2,.....,b_i), \binom{\lambda}{\lambda})=s_{i+1}$,

$\delta_i(s_j, \binom{\lambda}{\lambda})=s_{j+1}$, $i+1<=j<=n-1$,

$\delta_i(s_n, \binom{\lambda}{\lambda})=K_n$

In the $n^{th}$ component, its head reads the symbol $a_n$ and $b_n$, all other components are waiting and then it goes to state $q_j$, and simultaneously sends its current state information to all other components. Thus, all other components also go to state $q_j$.

$i=n$  $\delta_i(q_i, \binom{\lambda}{\lambda})=p_1$

$\delta_i(p_j, \binom{\lambda}{\lambda})=p_{j+1}$, $1<=j<=n-3$

$\delta_i(p_{n-2}, \binom{\lambda}{\lambda})=K_{n-1}$

$\delta_i((q_i, a_1,a_2,.....,a_{n-1},b_1,b_2,.....,b_{n-1}), \binom{a_n}{b_n})=(q_i, a_1,a_2,.....,a_n,b_1,b_2,.....,b_n)$,

$\delta_i((q_i, a_1,a_2,.....,a_n,b_1,b_2,.....,b_n), \binom{\lambda}{\lambda})=q_j$,

Thus, the parallel communicating Watson-Crick automata system simulates one transition of the multi-head Watson-Crick automaton in the above stated manner. Thus any transition multi-head Watson-Crick automaton makes, the parallel communicating Watson-Crick automata system can replicate it. Moreover if at a particular instance a multi-head Watson-Crick automaton does not have a transition defined and rejects the input, a similar behaviour is expected from the parallel communicating Watson-Crick automata system as at least one of its component will also not have transition defined as a result that component will halt before others and thus reject the input also.

**Theorem 6:** Non-deterministic multi-head Watson-Crick automata and parallel communicating Watson-Crick automata systems have the same computational power.

Proof: From Theorem 4 and Theorem 5, we see that for every parallel communicating Watson-Crick automata system there is a non-deterministic multi-head Watson-Crick automaton accepting the same language and vice-versa. Hence, we say that non-deterministic multi-head Watson-Crick automata and parallel communicating Watson-Crick automata systems have the same computational power.

**Lemma 5:** non-deterministic multi-head Watson-Crick automaton with 1+2 heads can accept the non-regular unary language $L=\{a^{n^2}, \text{where } n > 1\}$.

Proof: The A non-deterministic multi-head Watson-Crick automaton $M=(V,\rho,Q,q_0,F,\delta,k_1,k_2)$ accepts the unary non-regular language $L=\{a^{n^2}, \text{where } n > 1\}$ where $V=\{a,b,c\}$, $\rho=\{(a,b),(a,c)\}$, $Q=\{q_0, q_1, q_2, q_3, q_4, q_{1a}, q_{1aa}, q_{2a}, q_{3a}\}$, $F=\{q_4\}$, $k_1=1$, $k_2=2$.

The transitions of M are as follows:

$\delta(q_0,(\lambda|b,\lambda))=q_0$, $\delta(q_0,(\lambda|c,\lambda))=q_1$, $\delta(q_1,(a|\lambda,b))=q_{1a}$, $\delta(q_{1a},(a|\lambda,\lambda))=q_{1aa}$, $\delta(q_{1aa},(a|\lambda,\lambda))=q_2$, $\delta(q_2,(a|c,b))=q_2$, $\delta(q_2,(a|b,c))=q_{2a}$, $\delta(q_{2a},(a|\lambda,\lambda))=q_3$, $\delta(q_2,(\lambda|\lambda,b))=q_4$, $\delta(q_3,(a|b,c))=q_3$, $\delta(q_3,(a|c,b))=q_{3a}$, $\delta(q_{3a},(a|\lambda,\lambda))=q_2$, $\delta(q_3,(\lambda|\lambda,c))=q_4$, $\delta(q_4,(\lambda|\lambda,b))=q_4$, $\delta(q_4,(\lambda|\lambda,c))=q_4$.

The multi-head Watson-Crick automaton M works in the following manner:

We use the first lower head and the second lower head of M to check whether the lower strand have alternative blocks of b's and c's which are of equal size.

We employ the upper head and the first lower head of M to check whether the number of 'bc' and 'cb' pairs in the lower strand is one less than the total number of b's in the first block.

Multi-head Watson-Crick automaton begins in state $q_0$ and moves the first lower head until it comes across the first 'c' in the lower strand. When M comes across the first 'c' in the lower strand it goes to state $q_1$. If the lower strand is of the form $b^n c^n b^n c^n...$, then when the first lower head reads the first 'c' the upper head is n symbols behind the first lower head (as the first 'c' comes after n b's). When M comes across the first 'c' in its first lower head, its upper head reads three a's and its second lower head reads a 'b' and M goes to state $q_2$. After that, every time the first lower head of M reads a 'b'/'c' the upper head of M reads an 'a' and the second lower head of M reads a 'c/b'. As the second lower head of M was at the beginning of the tape when first lower head of M read the first 'c' if the input is of the form $b^n c^n b^n c^n...$ then for every 'b'/ 'c' read by the first lower head of M, the second lower head of M will read a 'c'/'b'. This ensures that the lower strand have alternate blocks of 'b's and 'c' s of equal size. If the lower strand does not have equal blocks of 'b' and 'c' then M will halt without both its heads reaching the end of the tape. Thus, M rejects the input.

Every time there is a change of symbol read by the first lower head of M from 'b' to 'c' or from 'c' to 'b' the upper head of M reads an extra 'a' in addition to the steps mentioned above. Thus, for each 'bc' or 'cb' pair except for the first 'bc' pair the upper head of M reads an extra 'a'. For the first 'bc' pair, the upper head of M reads two extra a's. The extra a's read by the upper head of M, for each 'bc' or 'cb' pair M comes across in its lower strand, enables the upper head of M which was n symbols behind the lower head to catch up with lower head and reach the end of the tape.

Therefore the upper head and first lower head of M reach the end of their respective tapes at the same time only when the number of 'bc' or 'cb' pairs is (n-1) and M goes to state $q_4$. If the total number of 'bc' and 'cb' pairs is more than (n-1) the upper head of M will reach the end of the tape earlier than first lower head as result when the first lower head of M reads a character in its lower strand there will be no 'a' available in the upper head to read thus the automaton M will halt with its upper head at the end of the tape and the first lower hand not at the end. If the total number of 'bc' and 'cb' pairs is less than (n-1) the first lower head will finish earlier than upper head and thus the automaton M will halt with its upper head still needing to consume 'a'. As there are no transitions in M where the upper head can read 'a' without the first lower head reading any character, thus upper head of M will not reach the end of the tape. Both the heads of M reach the end of their respective tapes only if the number of 'bc' or 'cb' pairs is one less than the size of the first block of 'b's. In state $q_4$, if the upper head and the first lower head are at the end of the tape, the second lower head consumes its input to reach its tape end and then M accepts the input as $q_4$ is a final state.

From the above explained structure of M, we see that the upper head and the two lower heads of M will reach the end of their respective tapes only when the upper strand of input to A contains only 'a's and lower strand is of the form $b^n c^n b^n c^n$... where the number of 'bc' and 'cb' pairs are (n-1).

Thus, if we analyze the string multi-head Watson-Crick automaton M accepts we will see that it accepts all those strings composed of letter 'a' which has a complementarity string in the form $b^n c^n b^n c^n$... where the total number of 'bc' and 'cb' pairs is (n-1).

Consider a string w in $L=\{a^{n^2}, \text{where } n > 1\}$, one of its many complementarity strings must be of the form $b^n c^n b^n c^n$... where the number of 'bc' and 'cb' pairs are (n-1). E.g. $a^{3^2}$ i.e. aaaaaaaaa where n=3 one of its many complementarity strings is bbbcccbbb. Therefore w is accepted by M.

Now consider a string w not in L, no matter what complementarity string of w we take it can never be of the form $b^n c^n b^n c^n$... where the number of 'bc' and 'cb' pairs are (n-1). So w will not be accepted by M.

Thus, we can say that M accepts L.

**Theorem 7:** Parallel communicating Watson-Crick automata system with just two components can accept the non-regular unary language $L=\{a^{n^2}, \text{where } n > 1\}$.

Proof: From Lemma 5, we see that non-deterministic multi-head Watson-Crick automaton with 1+2 heads can accept the non-regular unary language $L=\{a^{n^2}, \text{where } n > 1\}$ and from Theorem 5 we know that for every multi-head Watson-Crick automata with $k_1+k_2$ heads there is a parallel communicating Watson-Crick automata system with $\max(k_1,k_2)$ components which accepts the same language. Thus, there exists a parallel communicating Watson-Crick automata system with just two components that can accept the non-regular unary language $L=\{a^{n^2}, \text{where } n > 1\}$.

The above result improves on Czeizler et. al. work as Czeizler et. al. showed that parallel communicating Watson-Crick automata system can accept the non-regular unary language $L=\{a^{n^2}, \text{where } n > 1\}$ using three components.

### V. COMPUTATIONAL COMPLEXITY OF DETERMINISTIC MULTI-HEAD WATSON-CRICK AUTOMATA

In this Section, we discuss the computational power of deterministic multi-head Watson-Crick automata.

**Theorem 8:** Deterministic multi-head Watson-Crick automaton with 1+2 heads can accept the non-regular unary language $L=\{a^{n^2+1}, \text{where } n > 1\}$.

Proof: A deterministic multi-head Watson-Crick automaton $M=(V,\rho,Q,q_0,F,\delta,k_1,k_2)$ accepts the unary non-regular language $L=\{a^{n^2+1}, \text{where } n > 1\}$ where $V=\{a,b,c,\#\}$, $\rho = \{(a,b), (a,c), (a,\#)\}$, $Q=\{q_0, q_1, q_2, q_3, q_4, q_{1a}, q_{1aa}, q_{2a}, q_{3a}, q_{4b}, q_{4c}, q_5\}$, $F=\{q_5\}$, $k_1=1, k_2=2$.

The transitions of M are as follows:
$\delta(q_0,(\lambda|b,\lambda))=q_0$, $\delta(q_0,(\lambda|c,\lambda))=q_1$, $\delta(q_1,(a|\lambda,b))=q_{1a}$, $\delta(q_{1a},(a|\lambda,\lambda))=q_{1aa}$, $\delta(q_{1aa},(a|\lambda,\lambda))=q_2$, $\delta(q_2,(a|c,b))=q_2$, $\delta(q_2,(a|b,c))=q_{2a}$, $\delta(q_{2a},(a|\lambda,\lambda))=q_3$, $\delta(q_2,(a|\#,b))=q_4$, $\delta(q_3,(a|b,c))=q_3$, $\delta(q_3,(a|c,b))=q_{3a}$, $\delta(q_{3a},(a|\lambda,\lambda))=q_2$, $\delta(q_3,(a|\#,c))=q_4$, $\delta(q_4,(\lambda|\lambda,b))=q_4$, $\delta(q_4,(\lambda|\lambda,c))=q_4$, $\delta(q_4,(\lambda|\lambda,\#))=q_5$.

The multi-head Watson-Crick automaton M works in the following manner:

We use the first lower head and the second lower head of M to check whether the lower strand have alternative blocks of b's and c's which are of equal size.

We employ the upper head and the first lower head of M to check whether the number of 'bc' and 'cb' pairs in the lower strand is one less than the total number of b's in the first block.

Multi-head Watson-Crick automaton begins in state $q_0$ and moves the first lower head until it comes across the first 'c' in the lower strand. When M comes across the first 'c' in the lower strand it goes to state $q_1$. If the lower strand is of the form $b^n c^n b^n c^n...$, then when the first lower head reads the first 'c', the upper head is n symbols behind the first lower head (as the first 'c' comes after n b's). When M comes across the first 'c' in its first lower head, its upper head reads three a's and its second lower head reads a 'b' and M goes to state $q_2$. After that, every time the first lower head of M reads a 'b'/'c' the upper head of M reads an 'a' and the second lower head of M reads a 'c/b'. As the second lower head of M was at the beginning of the tape when first lower head of M read the first 'c' if the input is of the form $b^n c^n b^n c^n...$ then for every 'b'/ 'c' read by the first lower head of M, the second lower head of M will read a 'c'/'b'. This ensures that the lower strand have alternate blocks of 'b's and 'c' s of equal size. If the lower strand does not have equal blocks of 'b' and 'c' then M will halt without both its heads reaching the end of the tape. Thus, M rejects the input.

Every time there is a change of symbol read by the first lower head of M from 'b' to 'c' or from 'c' to 'b' the upper head of M reads an extra 'a' in addition to the steps mentioned above. Thus, for each 'bc'or 'cb' pair except for the first 'bc' pair the upper head of M reads an extra 'a'. For the first 'bc' pair, the upper head of M reads two extra a's. The extra a's read by the upper head of M, for each 'bc' or 'cb' pair M comes across in its lower strand, enables the upper head of M which was n symbols behind the lower head to catch up with lower head and reach the end of the tape.

Therefore the upper head and first lower head of M reach the end of their respective tapes at the same time only when the number of 'bc' or 'cb' pairs is (n-1) and M goes to state $q_4$(in the deterministic case unlike the non deterministic case explained before the end of the tape is identified by the symbol '#' in the lower strand). If the total number of 'bc' and 'cb' pairs is more than (n-1) the upper head of M will reach the end of the tape earlier than first lower head as result when the first lower head of M reads a character in its lower strand there will be no 'a' available in the upper head to read thus the automaton M will halt with its upper head at the end of the tape and the first lower hand not at the end. If the total number of 'bc' and 'cb' pairs is less than (n-1) the first lower head will finish earlier than upper head and thus the automaton M will halt with its upper head still needing to consume 'a'. As there are no transitions in M where the upper head can read 'a' without the first lower head reading any character, thus upper head of M will not reach the end of the tape. Both the heads of M reach the end of their respective tapes only if the number of 'bc' or 'cb' pairs is one less than the size of the first block of 'b's. In state $q_4$, if the upper head and the first lower head are at the end of the tape, the second lower head consumes its input to reach its tape end where its reads the symbol '#' and goes to state $q_5$, then M accepts the input as $q_5$ is a final state and all the heads are at the end of the tape.

From the above explained structure of M, we see that the upper head and the two lower heads of M will reach the end of their respective tapes only when the upper strand of input to A contains only 'a's and lower strand is of the form $b^n c^n b^n c^n...\#$ where the number of 'bc' and 'cb' pairs are (n-1).

Thus, if we analyze the string multi-head Watson-Crick automaton M accepts we will see that it accepts all those strings composed of letter 'a' which has a complementarity string in the form $b^n c^n b^n c^n...\#$ where the total number of 'bc' and 'cb' pairs is (n-1).

Consider a string w in L={$a^{n^2+1}$, where $n > 1$}, one of its many complementarity strings must be of the form $b^n c^n b^n c^n...\#$ where the number of 'bc' and 'cb' pairs are (n-1) for e.g. $a^{3^2}$ i.e. aaaaaaaaa where n=3 one of its many complementarity strings is bbbcccbbb#. Therefore w is accepted by M.

Now consider a string w not in L, no matter what complementarity string of w we take it can never be of the form $b^n c^n b^n c^n...\#$ where the number of 'bc' and 'cb' pairs are (n-1). So w will not be accepted by M.

Thus, we can say that M accepts L.

**Corollary 3:** A deterministic multi-head Watson-Crick automata with 1+2 heads accepts a language not accepted by any non-determinitic multi-head finite automata.

Proof: From Theorem 8, we know that a deterministic multi-head Watson-Crick automata with 1+2 heads can accept the non-regular unary language L={$a^{n^2+1}$, where $n > 1$} and we also know that non-deterministic multi-head finite automata can accept only regular unary languages[12]. Thus, deterministic multi-head Watson-Crick automata with 1+2 heads accepts a language not accepted by any non-deterministic multi-head finite automata.

**Theorem 9:** For every deterministic multi-head finite automaton with k+1 heads there is a deterministic multi-head Watson-Crick automaton with k+1 heads and identity complementarity relation.

Proof: The proof of this Theorem is due to the fact that the complemenatrity relation is the the identity relation therefore the content of the lower strand is same as the upper strand. Therefore deterministic multi-head Watson-Crick automaton with $k_1$=k and $k_2$=1 and identity complementarity relation just simulates the moves of non-deterministic multi-head finite automaton with k+1 heads.

**Theorem 10:** The language L = {#$w_1$*$x_1$.........#$w_n$*$x_n$\$|n≥0, $w_i$ ∈{a,b}*, $x_i$∈ {a,*b*}*, ∃i∃j :$w_i$=$w_j$, $x_i$≠$x_j$} is accepted by a deterministic multi-head Watson-Crick automaton with non-injective complementarity relation and 1+1 heads.

Proof: Let, M=(V,ρ,Q,$q_0$,F,δ,1,1) be a deterministic multi-head Watson-Crick automaton,
where V={a,b,$v_{m1}$,$v_{m2}$,#,*,\$},ρ={(a,a),(#,#),(#,$v_{m1}$),(#,$v_{m2}$),(b,b),(*,*),(\$,\$)}, Q ={$q_0$, $q_1$, $q_2$, $q_3$, $q_4$, $q_5$, $q_{5\$}$, $q_6$},F={ $q_6$},and we have the following transitions:

δ($q_0$,(#|#))=$q_0$, δ($q_0$,(a|a))=$q_0$, δ($q_0$,(b|b))=$q_0$, δ($q_0$,(*|*))=$q_0$, δ($q_0$,(#|$v_{m1}$))=$q_1$, δ($q_1$,(λ|a))=$q_1$, δ($q_1$,(λ|b))=$q_1$, δ($q_1$,(λ|*))=$q_1$, δ($q_1$,(λ|#))=$q_1$, δ($q_1$,(λ|$v_{m2}$))=$q_2$, δ($q_2$,(a|a))=$q_2$, δ($q_2$,(b|b))=$q_2$, δ($q_2$,(*|*))=$q_3$, δ($q_3$,(a|a))=$q_3$, δ($q_3$,(b|b))=$q_3$, δ($q_3$,(#|#))=$q_4$, δ($q_3$,(#|\$))=$q_4$, δ($q_3$,(a|b))=$q_5$, δ($q_3$,(a|*))=$q_5$, δ($q_3$,(a|#))=$q_5$, δ($q_3$,(a|\$))=$q_{5\$}$, δ($q_3$,(b|a))=$q_5$, δ($q_3$,(b|*))=$q_5$, δ($q_3$,(b|#))=$q_5$, δ($q_3$,(b|\$))=$q_{5\$}$, δ($q_3$,(*|a))=$q_5$, δ($q_3$,(*|b))=$q_5$, δ($q_3$,(*|#))=$q_5$, δ($q_3$,(*|\$))=$q_5$, δ($q_3$,(#|a))=$q_5$, δ($q_3$,(#|b))=$q_5$, δ($q_3$,(#|*))=$q_5$, δ($q_5$,(λ|x))=$q_5$ where x∈V-{\$}, δ($q_5$,(λ|\$))=$q_{5\$}$, δ($q_{5\$}$,(x|λ))=$q_{5\$}$ where x∈V-{\$} , δ($q_{5\$}$,(\$|λ))=$q_6$,

The above stated automaton works in the following manner:

The elements (%,$v_{m1}$) and (%, $v_{m2}$) of the complementarity relation ρ are used to guess the two substrings of the input string which has its w parts equal and x parts unequal. On finding these guessed substrings the automaton goes to state $q_2$. In state $q_2$, the automaton M checks to see whether the guessed substrings have their w parts equal or not. If the substrings do not have their w parts equal then the automaton halts in a non-final state as no transitions are defined for such a situation in state $q_2$ and the automaton rejects the input string. If the two guessed substring have their w parts equal then the automaton goes to state $q_3$. In state $q_3$, the automaton M checks whether the guessed substrings having their w parts equal have their x parts equal or not. If the x parts are equal then the automaton goes to state $q_4$. The state $q_4$ is a non-final state and no transitions are defined on state $q_4$, thus, the automaton rejects the input string. If the x parts are unequal then the automaton goes to state $q_5$ which signifies the guessed strings have their w parts equal but x parts unequal, therefore in $q_5$ the elements in the lower strand which are left is consumed until M comes across '\$' in the lower head then the control switches to $q_{5\$}$ where the elements in the upper strand is consumed until M comes across '\$' in the upper strand on encountering '\$' M goes to state $q_6$, both its head are at the end of their respective tapes and $q_6$ is a accepting state; thus the input string is accepted as the guessed substrings have their w parts equal and x parts unequal.

Consider a string s in L. One of the many complementarity strings of s will correctly guess the two substrings which have their w parts equal and x parts unequal and s will be accepted by the automaton M.

Now, consider a string s not in L. As s is not in L there are no two substrings of s whose w parts are equal and x parts are unequal. Therefore, no matter the guess made by any complementarity string of s for the location of two substrings of s they will never have their w parts equal and x parts unequal. So M rejects s. Thus, from the above stated arguments we conclude that M accepts L.

**Corollary 4:** A deterministic multi-head Watson-Crick automata with 1+1 heads accepts a language not accepted by any deterministic multi-head finite automata.

Proof: From Theorem 10, we know that a deterministic multi-head Watson-Crick automata can accept the language L = {#$w_1$*$x_1$.........#$w_n$*$x_n$\$|n≥0, $w_i$ ∈{a,b}*, $x_i$∈ {a,*b*}*, ∃i∃j :$w_i$=$w_j$, $x_i$≠$x_j$} with 1+1 heads. Moreover we also know that deterministic multi-head finite automata cannot accept L={#$w_1$*$x_1$.........#$w_n$*$x_n$\$|n≥0, $w_i$ ∈{a,b}*, $x_i$∈ {a,*b*}*, ∃i∃j :$w_i$=$w_j$, $x_i$≠$x_j$} [5]. Thus, deterministic multi-head Watson-Crick automata with 1+1 heads accepts a language not accepted by any deterministic multi-head finite automata.

**Theorem 11:** The set of languages accepted by deterministic multi-head finite automata is a proper subset of set of languages accepted by deterministic multi-head Watson-Crick automata.

Proof: From Theorem 9, we know that for every deterministic multi-head finite automaton there exists a deterministic multi-head Watson-Crick automaton which accepts the same language and from Corollary 4, we see that deterministic multi-head Watson-Crick automata accept a language which is not accepted by any deterministic multi-head finite automata. Thus, the set of languages accepted by deterministic multi-head finite automata is a proper subset of set of languages accepted by deterministic multi-head Watson-Crick automata

**Lemma 6:** For every deterministic parallel communicating Watson-Crick automata system A = (V, ρ, $A_1$, $A_2$, . . . , $A_n$, K), with n components there is a deterministic multi-head Watson-Crick automaton with n+n heads which accept the same language.

Proof: The proof of the above Lemma is same as the proof for Theorem 4. If the components of the parallel communicating Watson-crick automata are deterministic then the multi-head Watson-Crick automata constructed from the parallel communicating automata system is also deterministic.

**Lemma 7:** For every deterministic multi-head Watson-Crick automata with $k_1+k_2$ heads there is a deterministic parallel communicating Watson-Crick automata system with $max(k_1,k_2)$ components which accepts the same language.

Proof: The proof of the above Lemma is same as the proof for Theorem 5. If the multi-head Watson-Crick automata is deterministic then the components of the parallel communicating Watson-crick automata system constructed from the multi-head Watson-Crick automata is also deterministic.

**Theorem 12:** Deterministic multi-head Watson-Crick automata and deterministic parallel communicating Watson-Crick automata systems have the same computational power.

Proof: The proof follows from Lemma 6 and Lemma 7.

## VI. CONCLUSION

In this paper, we introduce a new model namely multi-head Watson-Crick automata. We discuss in details the computational power of both the deterministic and non-deterministic variant of this model. We establish the superiority of this model over multi-head finite automata and have also shown that both the deterministic and non-deterministic variant of this model can accept non-regular unary languages. Moreover we also compare this model with parallel communicating Watson-Crick automata system. The question whether the set of languages accepted by deterministic multi-head Watson-Crick automata is a proper subset of languages accepted by non-deterministic multi-head Watson-Crick automata is still unanswered.